\documentclass[12pt]{article}

\newlength{\spse}
\usepackage{amsfonts}
\usepackage{amsthm}
\input{epsf.sty}
\usepackage{latexsym,amsmath,amsfonts,amscd}
\usepackage{epsfig}
\usepackage{changebar}
\usepackage{pstricks}
\usepackage{pst-plot}
\usepackage{multirow}
\usepackage{subfigure}
\usepackage{placeins}
\usepackage{amssymb}
\usepackage{mathrsfs}
\usepackage[title]{appendix}
\usepackage{bm}
\usepackage{hyperref}

\usepackage{multirow,bigstrut}
\usepackage{enumitem}
\usepackage{booktabs}
\usepackage{xcolor}

\usepackage{chemformula}
\usepackage[linesnumbered,ruled,vlined]{algorithm2e}
\usepackage{mathtools, nccmath}

\DeclarePairedDelimiter{\nint}\lfloor\rceil

\SetCommentSty{mycommfont}

\numberwithin{equation}{section}

\theoremstyle{plain}
\newtheorem{thm}{Theorem}[section]

\newtheorem{rem}[thm]{Remark}

\theoremstyle{definition}
\newtheorem{exam}[thm]{Example}

\newcommand{\brac}[1]{\left(#1\right)}
\newcommand{\abs}[1]{\left\vert#1\right\vert}
\newcommand{\norm}[1]{\left\Vert#1\right\Vert}

\newfont{\iams}{msbm9}

\newcommand{\commentbis}[1]{}
\newcommand{\be}{\begin{eqnarray}}
\newcommand{\ee}{\end{eqnarray}}
\newcommand{\beno}{\begin{eqnarray*}}
\newcommand{\eeno}{\end{eqnarray*}}

\newcommand{\barr}[1]{\begin{array}{#1}}
\newcommand{\earr}{\end{array}}

\newcommand{\beq}{\begin{equation}}
\newcommand{\eeq}{\end{equation}}
\newcommand{\beqa}{\begin{eqnarray}}
\newcommand{\eeqa}{\end{eqnarray}}

\begin{document}
\baselineskip=2pc

\vspace{.5in}

\begin{center}

{\large\bf Data-driven discovery of multiscale chemical reactions \\ 
governed by the law of mass action}

\end{center}

\vspace{.1in}

\centerline{
Juntao Huang \footnote{Department of Mathematics, Michigan State University, East Lansing, MI 48824, USA. E-mail: huangj75@msu.edu} \quad
Yizhou Zhou \footnote{School of Mathematical Sciences, Peking University, Beijing, China. E-mail: zhouyz@math.pku.edu.cn}\quad
Wen-An Yong \footnote{Department of Mathematical Sciences, Tsinghua University, Beijing, China. E-mail: wayong@tsinghua.edu.cn.}
}

\vspace{.4in}

\centerline{\bf Abstract}

\vspace{.1in}

In this paper, we propose a data-driven method to discover multiscale chemical reactions governed by the law of mass action. 
First, we use a single matrix to represent the stoichiometric coefficients for both the reactants and products in a system without catalysis reactions. 
The negative entries in the matrix denote the stoichiometric coefficients for the reactants and the positive ones for the products.
Second, we find that the conventional optimization methods usually get stuck in the local minima and could not find the true solution in learning the multiscale chemical reactions. To overcome this difficulty, we propose a partial-parameters-freezing (PPF) technique to progressively determine the network parameters by using the fact that the stoichiometric coefficients are integers. With such a technique, the dimension of the searching space is gradually reduced in the training process and the global mimina can be eventually obtained. Several numerical experiments including the classical Michaelis–Menten kinetics, the hydrogen oxidation reactions and the simplified GRI-3.0 mechanism verify the good performance of our algorithm in learning the multiscale chemical reactions. The code is available at \url{https://github.com/JuntaoHuang/multiscale-chemical-reaction}.

\bigskip

\bigskip


\pagenumbering{arabic}

\section{Introduction}\label{sec:intro}
\setcounter{equation}{0}
\setcounter{figure}{0}
\setcounter{table}{0}

Chemical reactions are fundamental in many scientific fields including biology, material science, chemical engineering and so on. 
To identity the reactions from experimental data, the traditional methods are mainly based on some empirical laws and expert knowledge \cite{gao2016reaction}. Recently, thanks to the rapid development of machine learning \cite{lecun2015deep} and data-driven modeling \cite{rudy2017data,lusch2018deep,champion2019data,brunton2020machine,raissi2019physics,lu2019deepxde,raissi2020hidden,huang2020learning}, it is desirable to develop a data-driven method of discovering the underlying chemical reactions from massive data automatically.

Consider a reaction system with $n_s$ species participating in $n_r$ reactions:
$$
\nu_{i1}'\mathcal{S}_1 + \nu_{i2}'\mathcal{S}_2 + \cdots + \nu_{in_s}'\mathcal{S}_{n_s} \ch{<=>[$k\sb{if}$][$k\sb{ir}$]}
\nu_{i1}''\mathcal{S}_1 + \nu_{i2}''\mathcal{S}_2 + \cdots + \nu_{in_s}''\mathcal{S}_{n_s}
$$
for $i = 1,2, \cdots, n_r$. Here $\mathcal{S}_k$ is the chemical symbol for the $k$-th species, the nonnegative integers $\nu_{ik}'$ and $\nu_{ik}''$ are the stoichiometric coefficients of the $k$-th species in the $i$-th reaction, and $k_{if}$ and $k_{ir}$ are the direct and reverse reaction rates of the $i$-th reaction. The reaction is reversible if both $k_{if}$ and $k_{if}$ are positive. Strictly speaking, all elementary chemical reactions are reversible due to microscopic reversibility. However, in real applications, some of the rate constants are  negligible, thus the corresponding reactions can be omitted and the retained ones can be considered as irreversible.

Denote by $u_k = u_k(t)$  the concentration of the $k$-th species at time $t$ for $k=1,2,\cdots,n_s$. 
According to the law of mass action \cite{voit2015150}, the evolution of $u_k$ obeys the ordinary differential equations (ODEs) \cite{othmer2003analysis}
\begin{equation}\label{eq:reaction-ODE}
	\frac{du_k}{dt} = \sum_{i=1}^{n_r}(\nu_{ik}''-\nu_{ik}')\left(k_{if}\prod_{j=1}^{n_s}u_j^{\nu_{ij}'}-k_{ir}\prod_{j=1}^{n_s}u_j^{\nu_{ij}''} \right),
\end{equation}
for $k=1,2,\cdots,n_s$.
Given the concentration time series data $\{u_k(t_{n}), \ k=1,\cdots,n_s, \ n=1,\cdots,N \}$, our goal is to learn the stoichiometric coefficients $\nu_{ik}'$, $\nu_{ik}''$ and reaction rates $k_{if}$ and $k_{ir}$. 

In the literature there are already some works on this topic.
In \cite{burnham2008inference}, the authors applied linear regressions to infer the chemical reactions, with the assumption that the reactions are at most the result of bimolecular collisions and the total reaction order is not greater than two. In \cite{willis2016inference}, the linear regression was utilized with an L1 objective, which transforms the problem into a mixed-integer linear programming (MILP). This approach suffers from the same restrictive assumptions as in \cite{burnham2008inference}. In \cite{langary2019inference}, the authors presented an approach to infer the stoichiometric subspace of a chemical reaction network from steady-state concentration data profiles, which is then cast as a series of MILP.
In \cite{nagy2020automatic}, some chemically reasonable requirements were considered such as the mass conservation and the principle of detailed balance.
The deep neural networks (DNNs) were applied to extract the chemical reaction rate information in \cite{ranade2019ann,ranade2019extended}, but the weights are difficult to interpret physically.
In \cite{hoffmann2019reactive}, the authors adapted the sparse identification of nonlinear dynamics (SINDy) method \cite{brunton2016discovering,de2020pysindy} to the present problem. However, the approach relies on expert knowledge, which precludes the application in a new reaction system with unknown reaction pathways. 
Within the framework of SINDy, other works are \cite{Bhavana2019Machine,Bhavana2020Operable,mangan2016inferring}. In order to improve the performance of SINDy, two additional steps including least-squares regression and stepwise regression in the identification were introduced in \cite{Bhavana2019Machine}, which are based on the traditional statistical methods. In \cite{Bhavana2020Operable}, SINDy was combined with the DNNs to adaptively model and control the process dynamics. An implicit-SINDy was proposed and applied to infer the Michaelis-Menten enzyme kinetics in \cite{mangan2016inferring}.
Additionally, a statistical learning framework was proposed based on group-sparse regression which leverage prior knowledge from physical principles in \cite{maddu2020learning}. For example, the mass conservation is enforced in the JAK-STAT reaction pathway for signal transduction in \cite{maddu2020learning}.

{
Our work is mainly motivated by \cite{ji2020autonomous}, where the authors proposed a Chemical Reaction Neural Network (CRNN) by resorting to the feature of the equations in \eqref{eq:reaction-ODE}. The discovery of chemical reactions usually involves two steps: the identification of the reaction pathways (i.e., the stoichiometric coefficients) and the determination of the reaction rates. For complex reaction processes, one could not even identify the reaction pathways and has to infer both the stoichiometric coefficients and the rate constants from data. 
The work in \cite{ji2020autonomous} presents a neural network approach for discovering unknown reaction pathways from concentration data. The parameters in CRNN correspond to the stoichiometric coefficients and reaction rates and the network has only one hidden layer with the exponential activation functions. 
}

Different from CRNN in \cite{ji2020autonomous}, we use a single matrix of order $n_r\times n_s$ to represent the stoichiometric coefficients for both the forward and reverse reactions by assuming no catalysis reactions. The negative entries in the matrix denote the stoichiometric coefficients for the reactants and the positive for the products.  

On the other hand, the reaction rates often differ in a wide range of magnitudes, which causes a lot of troubles in learning the multiscale chemical reactions.
To provide some insights into this difficulty, we design a nonlinear regression problem to fit a polynomial with two terms, see \eqref{eq:regression-function} in Section \ref{sec:regression}. The given coefficients of the polynomial differ in several orders of magnitudes and the polynomial degree is to be determined. We find numerically that the conventional optimization algorithm usually gets stuck in the local minima and could not find the true solution. Another observation in the numerical experiment is that the learned polynomial degree of the terms with larger coefficient is close to the true solution. Inspired by this observation, we propose a partial-parameters-freezing (PPF) technique to escape from the local minima. Specifically, we perform a round operation on the learned polynomial degree which are close to integer in the optimization process if the loss function does not decrease. The revised algorithm works well for this problem. Some theoretical analysis is also provided to explain the numerical phenomenon.

We then generalize the PPF technique to learn the multiscale chemical reactions. Notice that the stoichiometric coefficients are integers. In the training process, if the loss function stops to decrease, the stoichiometric coefficients which are close to integers are rounded and then frozen afterwards. With such a treatment, the stoichiometric coefficients are gradually determined, the dimension of the searching space is  reduced in the training process, and eventually the global mimina can be obtained. Several numerical experiments including the classical Michaelis–Menten kinetics, the hydrogen oxidation reactions and the simplified GRI-3.0 mechanism verify that our method performs much better in learning the mutiscale chemical reactions.

This paper is organized as follows. In Section \ref{sec:regression}, we investigate a multiscale nonlinear regression problem numerically and theoretically. Our algorithm for learning the multiscale chemical reactions is presented in Section \ref{sec:method}. In Section \ref{sec:numerical}, the performance of the algorithm is validated through several numerical examples. Finally, conclusions and the outlook of future work are presented in Section \ref{sec:conclusion}.

\section{Multiscale nonlinear regression problem}\label{sec:regression}

To provide some insights into the difficulties in learning the multiscale chemical reactions, we consider a nonlinear regression problem to fit the following function:
\begin{equation}\label{eq:regression-function}
	y = f(x;\theta_1, \theta_2) = c_1 x^{\theta_1} + c_2 x^{\theta_2}.
\end{equation}
Here $c_1$ and $c_2$ are two given constants satisfying $\abs{c_1}\ll\abs{c_2}$, and $\theta_1, \theta_2$ are two integers to be determined. This simple toy model captures two key features of the multiscale chemical reactions. The first feature is that the right-hand side of the chemical reaction ODEs \eqref{eq:reaction-ODE} is polynomials and the stoichiometric coefficients are integers. The second one is that the multiscale chemical reactions often have reaction rates which differ in several orders of magnitudes.

Given the dataset $\{(x_i, y_i): ~ i=1,\cdots,N \}$, we define the loss function to be the mean squared error (MSE):
\begin{equation}\label{eq:regression-loss}
	\mathcal{L}(\theta_1, \theta_2) = \frac{1}{N}\sum_{i=1}^N(f(x_i; \theta_1, \theta_2) - y_i)^2,
\end{equation}
to estimate the parameters $\theta_1$ and $\theta_2$.
Next, conventional optimization methods can be used to obtain the estimation of $\theta_1$ and $\theta_2$.

In the numerical experiment, we take $c_1=1$ and $c_2=100$. The ground truth solutions are $\theta_1=1$ and $\theta_2=2$. The data $x_i$ for $i=1,\cdots,N$ are randomly sampled from a uniform distribution in $(0, 1)$ with the number of data $N=1000$, and $y_i=c_1x_i+ c_2x_i^2$. The Adam optimization method \cite{kingma2014adam} is applied with the full batch gradient decent. The learning rate is taken to be $10^{-4}$. The initial guess of $\theta_1$ and $\theta_2$ is randomly chosen in $(-1, 1)$.

\begin{figure}
    \centering
    \subfigure[loss vs. epoch]{
    \begin{minipage}[b]{0.48\textwidth}
    \includegraphics[width=1\textwidth]{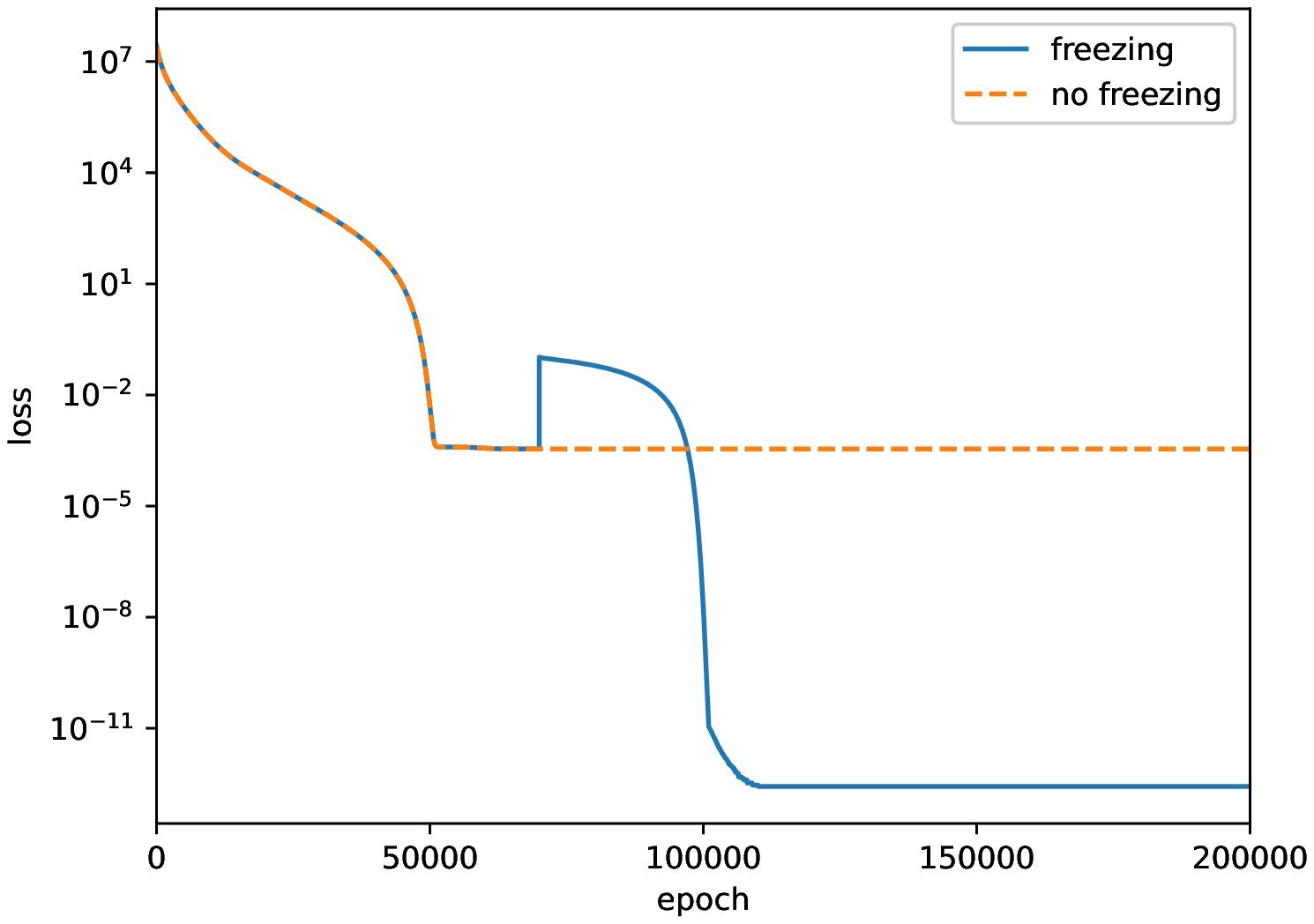}
    \end{minipage}
    }
    \subfigure[parameters $\theta_1$ and $\theta_2$ vs. epoch]{
    \begin{minipage}[b]{0.48\textwidth}    
    \includegraphics[width=1\textwidth]{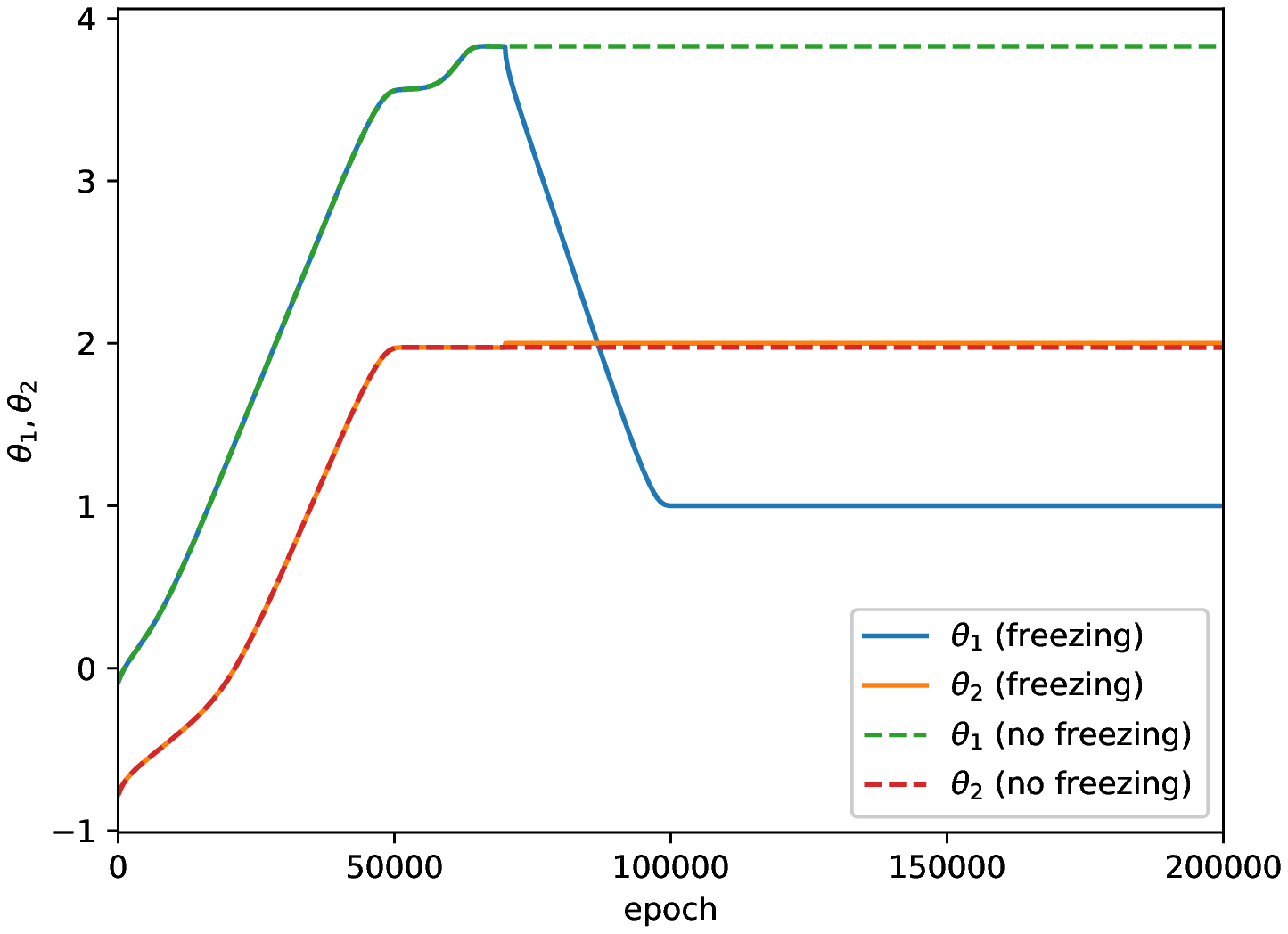}
    \end{minipage}
    }
    \caption{Multiscale nonlinear regression problem: the history of loss function in \eqref{eq:regression-loss} and the parameters $\theta_1$ and $\theta_2$ in the training process. Solid lines: the method with the PPF technique; dashed lines: the method without the PPF technique.}
    \label{regression:loss-history}
\end{figure}

For this toy model, we numerically find that the naive implementation will get stuck in the local minima $(\theta_1, \theta_2) = (3.8286, 1.9745)$ and could not find the true solution. The history of the loss function and the parameters $\theta_1$ and $\theta_2$ in the training is presented in Figure \ref{regression:loss-history}, see the dashed lines. 

Although the naive optimization could not find the global minima, we notice that $\theta_2=1.9745$ in this local minima is close to the true solution $\theta_2=2$. Inspired by this observation, we propose a partial-parameters-freezing (PPF) technique to escape from the local minima. To be more specific, we keep track of the loss function in the training. If the loss does not decrease, we check the parameters $\theta_1$ and $\theta_2$: if any of these is close to its nearest integer with a given threshold, we round it to the integer and do not update it in the afterwards optimization process.

For comparison, we also plot the history of the loss function and the parameters with the PPF technique in Figure \ref{regression:loss-history}, see the solid lines. The threshold is taken to be 0.05 in this test. The loss stops decreasing with the epoch around 7000. Then $\theta_2$ is rounded to 2 and only $\theta_1$ is updated afterwards. The true solution is eventually obtained when the epoch is around 10000.

\begin{figure}
    \centering
    \subfigure[loss function surface plot]{
    \begin{minipage}[b]{0.48\textwidth}
    \includegraphics[width=1.15\textwidth]{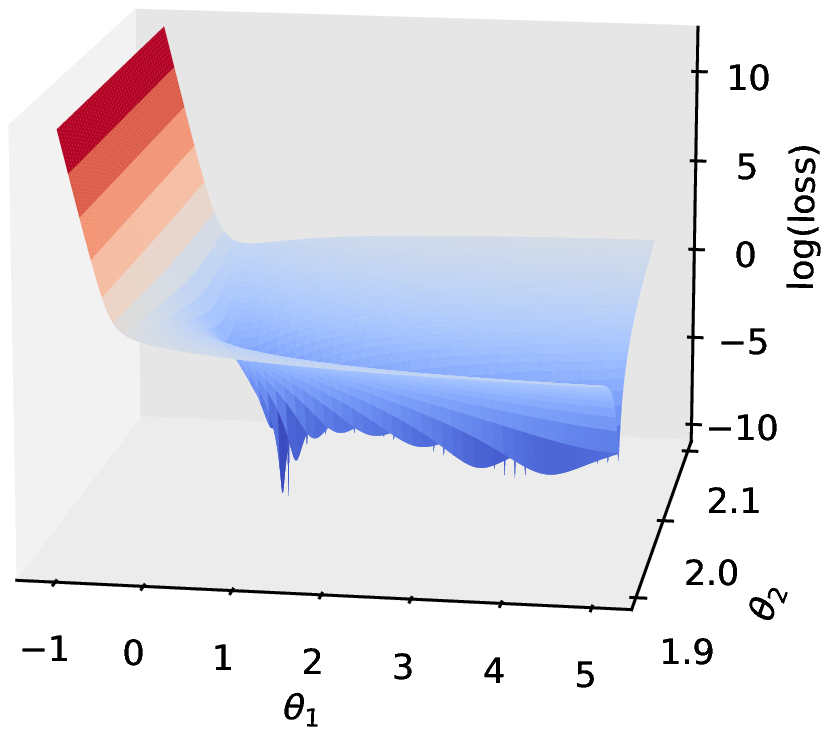}
    \end{minipage}
    }
    \subfigure[loss function contour map]{
    \begin{minipage}[b]{0.48\textwidth}    
    \includegraphics[width=1\textwidth]{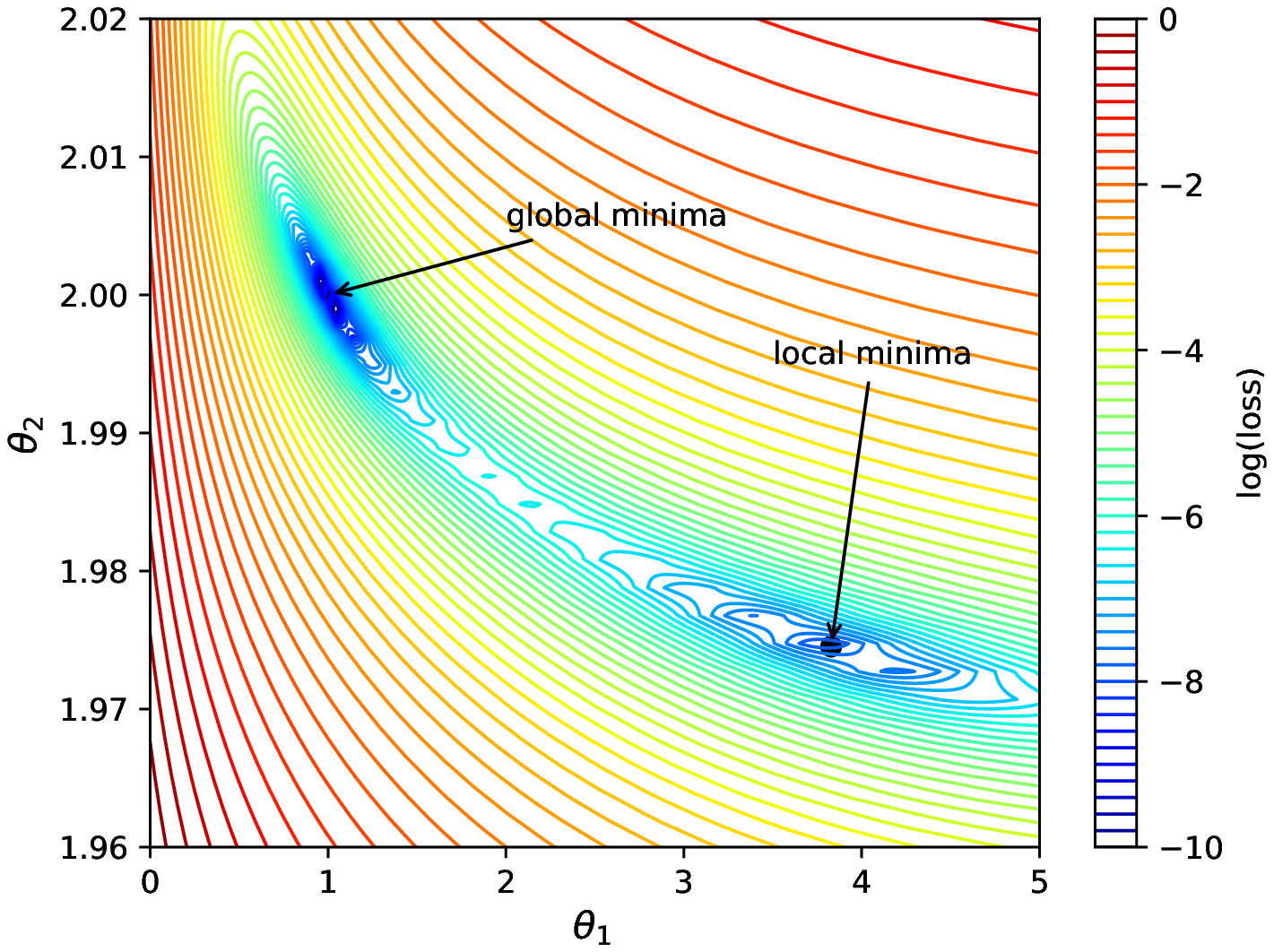}
    \end{minipage}
    }
    \caption{Multiscale nonlinear regression problem: the landscape of the loss function in \eqref{eq:regression-loss}. Left: loss function surface plot (in log scale); right: loss function contour map (in log scale), local minima $(\theta_1, \theta_2) = (3.8286, 1.9745)$.}
    \label{regression:loss-landscape}
\end{figure}

To better understand why it is easy to get stuck in the local minima without the PPF treatment, we investigate the landscape of the loss function. In Figure \ref{regression:loss-landscape}, we plot the 3D surface and the contour map for the loss as a function of $(\theta_1,\theta_2)$. In Figure \ref{regression:loss-landscape} (a), it is observed that the loss function has several local minima in which $\theta_2$ is close to 2. Moreover, the local minima $(\theta_1, \theta_2) = (3.8286, 1.9745)$ in the naive implementation is also labeled in Figure \ref{regression:loss-landscape} (b).

\begin{figure}
    \centering
    \includegraphics[width=0.5\textwidth]{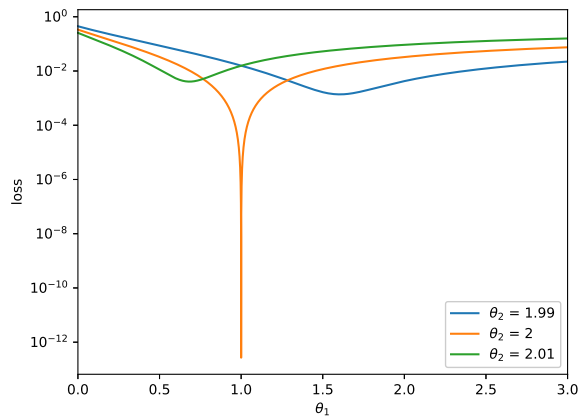}
    \caption{Multiscale nonlinear regression problem: loss function in \eqref{eq:regression-loss} with fixed parameters $\theta_2=1.99$, 2 and 2.01.}
    \label{regression:loss-1D}
\end{figure}

We also plot the profiles of the loss function with fixed $\theta_2=1.99$, 2 and 2.01 in Figure \ref{regression:loss-1D}. It is observed that slight perturbations in $\theta_2$ have a considerable impact on the minima of the loss function. Moreover, the loss as a 1D function with fixed $\theta_2=2$ is well-behaved. This explains why our algorithm is easy to find the global minima after freezing the integer parameter $\theta_2$.

We mention that we also test other cases with different coefficients $c_1$ and $c_2$ satisfying $\abs{{c_2}/{c_1}}=10^3, 10^4, 10^5$ and different integers $\theta_1$ and $\theta_2$. The results are similar and thus omitted here.

We conclude this section with some theoretical analysis to  explain the local minima phenomenon observed above. By taking gradient of the loss function in \eqref{eq:regression-loss}, we have
\begin{equation}\label{eq:regression-grad-loss}
	\frac{\partial\mathcal{L}}{\partial\theta_j} =  \frac{2c_jc_2 }{N}\sum_{i=1}^N  \brac{\frac{c_1}{c_2} (x_i^{\theta_1} - x_i^{\theta_1^{\textrm{e}}}) +  (x_i^{\theta_2} - x_i^{\theta_2^{\textrm{e}}})} x_i^{\theta_j} \ln x_i, \quad j=1,2.
\end{equation}
Here $\theta_i^{\textrm{e}}$ denotes the true solution of the parameter $\theta_i$ for $i=1,2$. 
From the expression \eqref{eq:regression-grad-loss}, we can provide some insights on the phenomenon that the local minina $\theta_2$ is close to the true solution $\theta_2^{\textrm{e}}$. 
To reach the local minima, the gradient should be zero. 
Refer to the expression \eqref{eq:regression-grad-loss}.
Since $\abs{c_1/c_2}\ll 1$, whether or not the gradient is close to zero depends mainly on the fact that $\theta_2$, instead of $\theta_1$, is close to the ground truth.

\section{Algorithm}\label{sec:method}

In this section, we present our algorithm for learning the multiscale chemical reactions. First, we use a single matrix to represent the stoichiometric coefficients for both the reactants and products. Each row of the matrix represents one reaction, where the negative entries denote the stoichiometric coefficients for the reactants and the positive ones for the products. This setup is valid for systems without catalysis reactions. In addition, we adapt the PPF technique for the multiscale nonlinear regression problem proposed in Section \ref{sec:regression} to learn the multiscale chemical reactions \eqref{eq:reaction-ODE}.

We assume that the data are given in the form of the concentrations and the time derivatives in different time snapshots $\{(u_k(t_{n}), u_k'(t_{n})), \ k=1,\cdots,n_s, \ n=1,\cdots,N \}$, our goal is to learn the stoichiometric coefficients and the reaction rates. Realistically, often only $u_k(t_{n})$ is available, and the time derivatives $u_k'(t_{n})$ could be approximated using numerical differentiations \cite{rudin1992nonlinear,chartrand2011numerical}.

To better illustrate the algorithm, we firstly introduce some vector notations. We denote the forward and reverse reaction rates in \eqref{eq:reaction-ODE} by $\bm{k}_f = (k_{1f},k_{2f},...,k_{n_rf})$ and $\bm{k}_r = (k_{1r},k_{2r},...,k_{n_rr})$. The stoichiometric coefficients in \eqref{eq:reaction-ODE} are collected in two matrices:
\begin{equation}
	\bm{V}' = 
	\begin{pmatrix}
	\nu_{11}' & \nu_{12}' & \cdots & \nu_{1n_s}'\\
	\nu_{21}' & \nu_{22}' & \cdots & \nu_{2n_s}'\\
	\vdots & \vdots & \ddots & \vdots\\
	\nu_{n_r1}' & \nu_{n_r2}' & \cdots & \nu_{n_rn_s}' 
	\end{pmatrix}
	, \qquad 
	\bm{V}'' = 
	\begin{pmatrix}
	\nu_{11}'' & \nu_{12}'' & \cdots & \nu_{1n_s}''\\
	\nu_{21}'' & \nu_{22}'' & \cdots & \nu_{2n_s}''\\
	\vdots & \vdots & \ddots & \vdots\\
	\nu_{n_r1}'' & \nu_{n_r2}'' & \cdots & \nu_{n_rn_s}'' 
	\end{pmatrix}.
\end{equation}

Assume that there is no catalysis reactions. Therefore, only one of $\nu_{ik}'$ and $\nu_{ik}''$ can be non-zero for any $(i,k)$.
In this case, the matrix $\bm{V}=(\nu_{ik}):=\bm{V}''-\bm{V}'$ satisfies
$$
\nu_{ik}=\left\{ {\begin{array}{*{20}c}
\vspace{1.5mm} \nu_{ik}'',\qquad & \textrm{if} \quad \nu_{ik}\ge0,\\[2mm]
               -\nu_{ik}',\qquad & \textrm{if} \quad \nu_{ik}<0.
\end{array}} \right.
$$
According to this property, we only need to pin down the matrix $\bm{V}$. Then $\bm{V}'$ and $\bm{V}''$ can be recovered by $\nu_{ik}''=\max(0,\nu_{ik})$ and $\nu_{ik}'=-\min(0,\nu_{ik})$, respectively.

Next, we define the neural network $\mathcal{N}=\mathcal{N}(u_1,\cdots,u_{n_s}):\mathbb{R}^{n_s}\rightarrow\mathbb{R}^{n_s}$ which has the input $\bm{u}:=(u_1,\cdots,u_{n_s})$ and the parameters $\bm{l}_f = (l_{1f}, l_{2f},..., l_{n_rf})$, $\bm{l}_r = (l_{1r}, l_{2r},..., l_{n_rr})$ and $\bm{V}$:
\begin{equation*}
	\mathcal{N}(u_1,\cdots,u_{n_s})_k = \sum_{i=1}^{n_r}\nu_{ik}\left(\exp({l_{if}})\prod_{j=1}^{n_s}u_j^{-\min(0,\nu_{ik})} - \exp({l_{ir}})\prod_{j=1}^{n_s}u_j^{\max(0,\nu_{ik})} \right)
\end{equation*}
for $k=1,\cdots,n_s$.
Here the parameters $l_{if}$ and $l_{ir}$ denote the logarithms of the reaction rates $k_{if}$ and $k_{ir}$ \cite{ji2020autonomous}. This change of variables technique has two advantages. The first one is that the positivity of the reaction rates is guaranteed automatically. The second one is that the reaction rates for the multiscale chemical reactions usually differ in several orders of magnitudes. The slight changes of $l_{if}$ and $l_{ir}$ will make $k_{if}$ and $k_{ir}$ change a lot, which could potentially make the neural network to be more robust in the training process.

The loss function is defined as the mean squared error (MSE) between the data for the time derivatives and the output of the neural network:
\begin{equation}\label{3.2}
	\mathcal{L} =  \frac{1}{N}\sum_{n=1}^N\sum_{k=1}^{n_s}\brac{\mathcal{N}(u_1(t_{n}),\cdots,u_{n_s}(t_{n}))_k - u_k'(t_{n})}^2 + \lambda  \mathcal{L}_r.	
\end{equation}
Here $\lambda  \mathcal{L}_r$ is a regularization term with $\lambda>0$ the regularization constant and
\begin{equation}
	\mathcal{L}_r = \sum_{i=1}^{n_r}\sum_{k=1}^{n_s} \abs{\nu_{ik}} + \sum_{i=1}^{n_r}\sum_{k=1}^{n_s} \nu_{ik}^2 + \sum_{i=1}^{n_r}(\abs{l_{if}} + \abs{l_{ir}}) + \sum_{i=1}^{n_r}(l_{if}^2 + l_{ir}^2).
\end{equation}
Here both $L_1$ and $L_2$ regularization terms are included.

This neural network works quite well for non-stiff chemical reactions. However, for stiff reactions, we observe that the optimization usually gets stuck in the local minima in the training process and could not find the true solution. The common techniques such as the mini-batch and reducing the learning rate do not work in such a situation. To attack this problem, we adapt the PPF technique proposed in the previous section.

The training procedure is split into two parts. The first part is to learn the matrix $\bm{V}$. To better illustrate the algorithm, we introduce some notation. Denote the vector in the $j$-th row of $\bm{V}$ by $\bm{v}_j$ for $j=1,\cdots,n_r$. Define the distance to the nearest integer for any vector $\bm{v}\in\mathbb{R}^{n_s}$ as
\begin{equation}\label{3.4}
	d_{\textrm{int}}(\bm{v}) := \norm{\bm{v} - \nint{\bm{v}}}_{\infty} = \max_{i\in\{1,\cdots,n_s\}} \abs{v_i - \nint{v_i}},
\end{equation}
where $\nint{}$ denotes the function rounding an arbitrary real number to its nearest integer and it is defined to work element-wise on vectors. We keep track of the loss function in the training process. If the loss function stops decreasing, we check if any row of $\bm{V}$ is close to the nearest integers, i.e., $d_{\textrm{int}}(\bm{v}_j)\le\epsilon$. Here, $\epsilon>0$ is a hyperparameter and we take $\epsilon=0.05$ in all the numerical examples in Section \ref{sec:numerical}. If the $j$-th row of $\bm{V}$ satisfies the condition $d_{\textrm{int}}(\bm{v}_j)\le\epsilon$, then we round $\bm{v}_j$ to $\nint{\bm{v}_j}$ and do not update it in the afterwards training. In addition, to help the optimization algorithm escape from the local minima, we randomly reinitialize other non-integer entries in $\bm{V}$ when the loss stops decreasing. After all the entries in $\bm{V}$ reach integer, we freeze them and then learn the parameters $\bm{l}_f$ and $\bm{l}_r$ related to the reaction rates. We remark that the SINDy algorithms \cite{brunton2016discovering,hoffmann2019reactive} can also be applied in learning the reaction rates when the stoichiometric coefficients $\bm{V}$ are known. The algorithm is summarized in Algorithm \ref{algorithm:stiff}.

\begin{rem}
	Here we assume that all the reactions are reversible. However, the algorithm can be also applied to irreversible reactions without any modification. The expected result is that the learned reverse reaction rates for the irreversible reactions will be close to zero. This will be demonstrated numerically in Example \ref{exam:enzyme} in Section \ref{sec:numerical}.
\end{rem}

{
\begin{rem}
	The number of reactions can be learned by repeatedly executing the algorithm with different $n_r$. The ground truth of $n_r$ can be inferred from the best one. This will be shown in the numerical examples in the next section.
\end{rem}
}

{

\begin{rem}
In many chemical reaction systems, the rate constants usually depend on the temperature. For example, the Arrhenius law can describe such a dependence:
	\begin{equation}
		k = A \exp\brac{-\frac{E_a}{RT}},
	\end{equation}
	where $k$ is the reaction rate, $A$ is the pre-exponential factor, $E_a$ is the activation energy and $R$ is the gas constant. In this case, the unknown parameters will include the pre-exponential factor, the activation energy and the stoichiometric coefficients. Our PPF technique can be directly applied without much modification. The performance will be verified numerically in the test in the next section.
\end{rem}
}

\begin{algorithm}[H]
\SetAlgoLined
\SetKwInOut{Input}{Input}\SetKwInOut{Output}{Output}

\Input{time series data $\{(u_k(t_{n}), u_k'(t_{n})), \ k=1,\cdots,n_s, \ n=1,\cdots,N \}$}
\Output{stoichiometric coefficient matrix $\bm{V}$, chemical reaction rates $\bm{k}_f$ and $\bm{k}_r$}

initialize hyperparameters: number of reactions $n_r$, total number of epoch $M$, learning rate $lr$, regularization coefficient $\lambda$, integer threshold $\epsilon$ \;

initialize parameters: $\bm{V}$, $\bm{l}_f$ and $\bm{l}_r$\;

\tcp{step 1: learning ${V}$}
$L_{\textrm{rec}}$ = np.zeros($M$);  \tcp*[f]{record loss function in each epoch} \\
$S_{\textrm{int}}=[~]$; 

\For{$i=1,\cdots,M$}
{	
	Compute loss $\mathcal{L}$ \;
 	
 	Compute $\frac{\partial \mathcal{L}}{\partial\theta}$ by backpropagation \;
 	
 	Update parameters (excluding the integer entries in $\bm{V}$) by Adam method \;

 	\tcp{if loss increase, then check if any row of ${V}$ is close to integer}
	\If{$L_{rec}[i] \ge L_{rec}[i-1]$}
	{
	    \For{$j=1,\cdots,n_r$}
	    {
	    	\If{$d_{\textrm{int}}(\bm{v}_j)\le\epsilon$}
	    	{
	    		$S_{\textrm{int}}$.append(j)\;
	    		$\bm{v}_j \leftarrow \nint{\bm{v}_j} $ \;
	    	}
	    	\If{j $\notin$ $S_{\textrm{int}}$}
	    	{
	    		$\bm{v}_j \leftarrow \textrm{rand}(-2, 2)$;		\tcp*[f]{random reinitialize non-integer entries in $V$} \\
	    	}	    	
	    }
	}

	\tcp{if all the entries in $V$ are integers, then stop learning $V$}
	\If{$S_{\textrm{int}} = \{ 1,\cdots,n_r \}$}
	{
		\textbf{break}\;
	}
} 

\tcp{step 2: learning $k_f$ and $k_r$}
\For{$i=1,\cdots,M$}
{
	Compute loss $\mathcal{L}$ \;
 	
 	Compute $\frac{\partial \mathcal{L}}{\partial\theta}$ by backpropagation \;
 	
 	Update parameters $\theta$ (excluding $\bm{V}$) by Adam method \;
}
\For{$i=1,\cdots,n_r$}
{
	$k_{if} \leftarrow \exp({l_{if}})$ \;
	$k_{ir} \leftarrow \exp({l_{ir}})$ \;	
}
\caption{Algorithm for learning chemical reactions}
\label{algorithm:stiff}
\end{algorithm}

\section{Numerical results}\label{sec:numerical}

Here the performance of our algorithm will be shown with five examples. The first example is an artificial reaction mechanism with two reactions \cite{lu2006applicability}. The second one is the well-known Michaelis-Menten kinetics \cite{keener1998mathematical} in biochemistry. The third one is the hydrogen oxidation reactions \cite{gorban2005invariant,chiavazzo2008quasi}. The fourth one is the extended Zeldovich mechanism, a typical chemical mechanism describing the oxidation of nitrogen and \ch{NOx} formation \cite{zeldovich1985mathematical}. The last one is the simplified GRI-3.0 mechanism, a chemical mechanism describing the methane oxidation \cite{2001Augmented}.

In each numerical example, we randomly take 100 different initial conditions to generate the data. For each initial condition, we take uniform time snapshots at $t_n=n\Delta t$ with $n=0,\dots,10$ and $\Delta t=0.1$. The data is generated by solving the governing ODEs numerically using implicit Runge-Kutta method of the Radau IIA family of the fifth order \cite{wanner1996solving} with small enough tolerance. The datasets are randomly split into the training datasets and the validation datasets by a ratio of 4:1. It is worthy to note that here we do not take $\Delta t$ to be too small so that the datasets could be potentially replaced by the experiment data in the future. The algorithm is implemented with PyTorch \cite{paszke2019pytorch}.

{ Now we present some details of the training and validation for the following four numerical tests. In the training process, all the parameters in the neural network are first randomly initialized from the uniform distribution in the interval $(-0.5, 0.5)$. Then, we update the parameters by minimizing the loss in \eqref{3.2} using the standard Adams algorithm \cite{kingma2014adam}. The learning rate is taken to be $10^{-3}$ and the regularization coefficient $\lambda$ in \eqref{3.2} is $10^{-8}$. Recall the training method following \eqref{3.4}, we take the integer threshold to be 0.05. Besides, the total epoch number is $10^6$ and the mini-batch gradient descent is applied with the batch size 10. 
For the validation, we use the following relative $L^2$ error:
\begin{equation*}
	E =  \sqrt{\frac{\sum_{n=1}^N\sum_{k=1}^{n_s}\abs{\mathcal{N}(u_1(t_{n}),\cdots,u_{n_s}(t_{n}))_k - u_k'(t_{n})}^2}{\sum_{n=1}^N\sum_{k=1}^{n_s}\abs{ u_k'(t_{n})}^2}}.	
\end{equation*}
Here the $(u_k(t_{n}), u_k'(t_{n}))$'s 
come from the validation dataset.
For the other details, we refer the interested readers to our code in \url{https://github.com/JuntaoHuang/multiscale-chemical-reaction}}.

\begin{exam}[hypothetical stiff reaction network]\label{exam:hypothetical}
The first test case is an artificial reaction network with two reactions, taken from \cite{lu2006applicability}:
\begin{subequations}\label{eq:artificial-reaction}
\begin{align}
	\ch{F <=>[$k\sb{1}\sp{+}$][$k\sb{1}\sp{-}$] R} \label{eq:artificial-reaction-1} \\
	\ch{R <=>[$k\sb{2}\sp{+}$][$k\sb{2}\sp{-}$] P} \label{eq:artificial-reaction-2}
\end{align}
\end{subequations}
Here F, R and P indicate the fuel, radical and product in combustions, respectively. The reaction rates are taken to be $k_1^+ = k_1^- = 1$ and $k_2^+ = k_2^- = 10^3$. The two reversible reactions in \eqref{eq:artificial-reaction} have dramatically different reaction rates. Thus, the second reaction \eqref{eq:artificial-reaction-2} will quickly approach to equilibrium after a transient period, after which the first one \eqref{eq:artificial-reaction-1} becomes rate-limiting. {This simple model is chosen to test the correctness of our code for stiff reactions.}

The corresponding ODE system for \eqref{eq:artificial-reaction} is linear. The eigenvalues of the coefficient matrix are $\lambda_1=-2000$, $\lambda_2=-1.5$ and $\lambda_3=0$, which differ in several orders of magnitudes. This indicates that the ODE system is stiff \cite{wanner1996solving}.

To illustrate the advantage of the PPF technique, we compare the performance of the algorithm with and without this technique. The history of the training and validation errors is shown in Figure \ref{fig:hypothetical-loss-freeze}. The relative error stays around $10^{-3}$ without this technique, and decreases to $10^{-6}$ after applying this technique. The learned parameters are listed in Table \ref{tab:hypothetical-params-freeze}. The upper part of the table is the learned parameters with the PPF technique, which agrees well with the ground truth in \eqref{eq:artificial-reaction}. By contrast, the algorithm without imposing this technique could not generate the correct result. Moreover, it is interesting to see that, without using the technique, the learned stoichiometric coefficients in the first reaction and the second one has the opposite sign. We also notice that the summation of the forward rate $k_f$ of the first reaction and the reverse rate $k_r$ of the second one is close to the true reaction rate $10^3$. The same holds true for the reverse rate of the first reaction and the forward rate of the second one. This indicates that the effect of these two learned reactions is identical to the fast reaction \eqref{eq:artificial-reaction-2} and the slow reaction \eqref{eq:artificial-reaction-1} is not captured here. This is similar to the phenomenon we observed in the multiscale nonlinear regression problem in Section \ref{sec:regression}.
\begin{figure}
    \centering
    \includegraphics[width=0.5\textwidth]{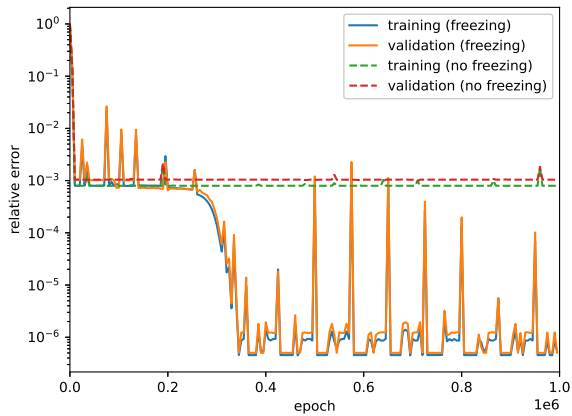}
    \caption{Example \ref{exam:hypothetical}: the history of the relative error for the training data and the verification data. Solid line: the method with the PPF technique; dashed line: the method without the PPF technique.}
    \label{fig:hypothetical-loss-freeze}
\end{figure}

\begin{table}[htbp]
  \centering
 \begin{tabular}{c|c|c|c|c|c}
    \hline
   freezing & $x_1$ & $x_2$ & $x_3$ & $k_f$ & $k_r$\\
    \hline
    $1$ 	& $0.000$  & $1.000$ & $-1.000$  & 1.000e+03 & 1.000e+03 \\
    $2$ 	& $-1.000$  & $1.000$  & $0.000$ & 1.000e+00 & 1.000e+00 \\
    \hline
   no freezing & $x_1$ & $x_2$ & $x_3$ & $k_f$ & $k_r$\\
    \hline
    $1$ 	& $-0.001$  & $0.999$ & $-0.999$  & 7.448e+02 & 5.731e+02 \\
    $2$ 	& $0.000$  & $-0.999$  & $0.999$ &  4.277e+02 & 2.559e+02 \\
    \hline    
    \end{tabular}
     \caption{Example \ref{exam:hypothetical}: learned parameters. Upper part: with the PPF technique; lower part: without the PPF technique. Here $(x_1, x_2, x_3)$ denotes the row vector of the matrix $\bm{V}$.}
     \label{tab:hypothetical-params-freeze}
\end{table}

Next, we test the algorithm with different number of chemical reactions. We take the number of reactions ranging from 1 to 4. The relative errors in the training data and the validation data are shown in Figure \ref{fig:hypothetical-loss-nodes}. The relative error decreases by three magnitudes when increasing the number of proposed reactions from one to two and reaches a plateau after that. Moreover, it is observed from Table \ref{tab:hypothetical-params-nodes} that some of the learned stoichiometric coefficients or reaction rates are close to zero if the number of reactions are larger than two. It then can be inferred that the kinetics could be well described with two reactions. 
\begin{table}[htbp]
  \centering
 \begin{tabular}{c|c|c|c|c|c}
    \hline
   reaction num 1 & $x_1$ & $x_2$ & $x_3$ & $k_f$ & $k_r$\\
    \hline
    $1$ 	& $0.000$  & $-1.000$ & $1.000$  & 1.000e+03 & 1.000e+03 \\
    \hline
   reaction num 2 & $x_1$ & $x_2$ & $x_3$ & $k_f$ & $k_r$\\    
   \hline
    $1$ 	& $0.000$  & $1.000$ & $-1.000$  & 1.000e+03 & 1.000e+03 \\
    $2$ 	& $-1.000$  & $1.000$  & $0.000$ & 1.000e+00 & 1.000e+00 \\    
    \hline
   reaction num 3 & $x_1$ & $x_2$ & $x_3$ & $k_f$ & $k_r$\\
    \hline
    $1$ 	& $0.000$  & $1.000$ & $-1.000$  & 1.000e+03 & 1.000e+03 \\
    $2$ 	& $-1.000$  & $1.000$  & $0.000$ & 9.217e+02 & 1.218e+02 \\
    $3$ 	& $0.750$  & $-0.101$  & $0.384$ & 7.887e$-$04 & 1.813e$-$03 \\
    \hline
   reaction num 4 & $x_1$ & $x_2$ & $x_3$ & $k_f$ & $k_r$\\
    \hline
    $1$ 	& $0.000$  & $-1.000$ & $1.000$  & 1.000e+03 & 1.000e+03 \\
    $2$ 	& $0.000$  & $0.000$  & $0.000$ & 5.931e+01 & 5.929e+01 \\
    $3$ 	& $0.000$  & $0.000$  & $0.000$ & 2.526e+01 & 2.524e+01 \\
    $4$ 	& $1.000$  & $-1.000$  & $0.000$ & 1.000e+00 & 1.000e+00 \\
	\hline
    \end{tabular}
     \caption{Example \ref{exam:hypothetical}: learned parameters with different number of reactions. Here $(x_1, x_2, x_3)$ denotes the row vector of the matrix $\bm{V}$.}
     \label{tab:hypothetical-params-nodes}
\end{table}

\begin{figure}
    \centering
    \includegraphics[width=0.5\textwidth]{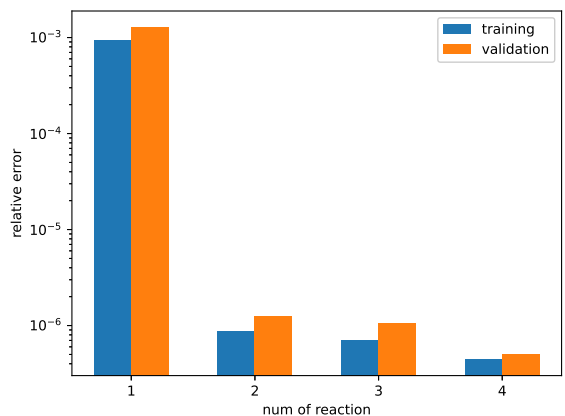}
    \caption{Example \ref{exam:hypothetical}: relative error for the training data and the validation data with different number of reactions.}
    \label{fig:hypothetical-loss-nodes}
\end{figure}

\end{exam}

\begin{exam}[enzyme kinetics]\label{exam:enzyme}

In this example, we consider the Michaelis–Menten kinetics \cite{keener1998mathematical}, one of the best-known models of enzyme kinetics in biochemistry. It involves an enzyme E, binding to a substrate S, to form a complex ES, which in turn releases a product P, regenerating the original enzyme. This can be represented schematically as \cite{keener1998mathematical}
\begin{equation}\label{eq:enzyme-reaction}
	\ch{E + S {<=>[$k\sb{f}$][$k\sb{r}$]} ES {->[$k\sb{cat}$]} E + P}
\end{equation}
Here $k_f$ denotes the forward rate constant, $k_r$ the reverse rate constant, and $k_{cat}$ the catalytic rate constant. This model is used in a variety of biochemical situations other than enzyme-substrate interaction, including antigen–antibody binding, DNA-DNA hybridization, and protein–protein interaction \cite{nelson2008lehninger}. Moreover, the reaction rates vary widely between different enzymes. In our test case, we follow \cite{srinivasan1986stage} and take $k_f=10^6$, $k_r=10^3$ and $k_{cat}=10$.

{Note that the second reaction in \eqref{eq:enzyme-reaction} is not reversible.} Here, we show that the exactly same algorithm can be applied to this situation. The results with and without the PPF technique are listed in Table \ref{tab:enzyme-params-freeze}. In the upper part of the table, the reverse rate for the second reaction is $1.949\times10^{-4}$. It then can be inferred that the system could be well described using two reactions with the second one to be irreversible. Again, the algorithm without this treatment could only get the correct result for the first faster reaction in \eqref{eq:enzyme-reaction}. The evolution of the loss function is similar to that in Example \ref{exam:hypothetical} and thus omitted here.
\begin{table}[htbp]
  \centering
 \begin{tabular}{c|c|c|c|c|c|c}
    \hline
   freezing & $x_1$ & $x_2$ & $x_3$ & $x_4$ & $k_f$ & $k_r$\\
    \hline
    $1$ 	& $-1.000$  & $-1.000$ & $1.000$ & $0.000$ & 1.000e+06 & 1.000e+03 \\
    $2$ 	& $1.000$  & $0.000$ &  $-1.000$ & $1.000$ & 1.000e+01 & 1.949e-04 \\
    \hline
   no freezing & $x_1$ & $x_2$ & $x_3$ & $x_4$ & $k_f$ & $k_r$\\
    \hline
    $1$ 	& $-0.999$  & $-0.999$ & $0.989$ & $0.000$ & 9.921e+05 & 9.929e+02 \\
    $2$ 	& $-1.001$  & $-1.000$ & $2.385$ & $0.000$ & 7.956e+03 & 1.392e+01 \\
    \hline    
    \end{tabular}
     \caption{Example \ref{exam:enzyme}: learned parameters. Upper part: with the PPF technique; lower part: without the PPF technique. Here $(x_1, x_2, x_3, x_4)$ denotes the row vector of the matrix $\bm{V}$.}
     \label{tab:enzyme-params-freeze}
\end{table}

{
{Next, we test the performance of the algorithm when the reaction rates depend on temperature.} We assume that the rate constants in \eqref{eq:enzyme-reaction} satisfy the Arrhenius law:
\begin{equation}
	k_f = A_f \exp\brac{-\frac{E_{a,f}}{R T}}, \quad k_r = A_r \exp\brac{-\frac{E_{a,r}}{R T}}, \quad k_{cat} = A_{cat} \exp\brac{-\frac{E_{a,cat}}{R T}}
\end{equation}
where the pre-exponential factors are given by
\begin{equation}
	A_f = 1, \quad A_r = 4, \quad A_{cat} = 10^3
\end{equation}
and the activation energy are
\begin{equation}
	E_f = 1600, \quad E_r = 3680, \quad E_{cat} = 2240
\end{equation}
and the gas constant $R = 8.3145$. The temperature is randomly taken in a uniform distribution in the interval $[200, 400]$. {In this case, the unknown parameters will include the pre-exponential factor, the activation energy in the Arrhenius law, and the stoichiometric coefficients. Our PPF technique can be directly applied without much modification.}

We compare the performance of the algorithm with and without the PPF technique. The history of the relative error for the training data and the verification data with variable temperature is shown in Figure \ref{fig:variable-temperature-loss-freeze}. We see clearly that the errors with the PPF technique are much smaller than those without the technique. We also show the learned parameters in Table \ref{tab:enzyme-params-freeze-temperature}. The upper part of the table is the learned parameters with the PPF technique, which agrees well with the ground truth. By contrast, the algorithm without imposing this technique could not generate the correct result. 
\begin{figure}
    \centering
    \includegraphics[width=0.5\textwidth]{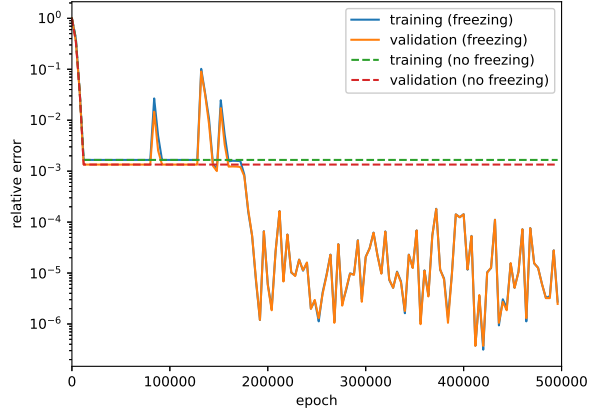}
    \caption{{Example \ref{exam:enzyme}: the history of the relative error for the training data and the verification data with variable temperature. Solid line: the method with the PPF technique; dashed line: the method without the PPF technique.}}
    \label{fig:variable-temperature-loss-freeze}
\end{figure}
\begin{table}[htbp]
  \centering
 \begin{tabular}{c|c|c|c|c|c|c|c|c}
    \hline
   freezing & $x_1$ & $x_2$ & $x_3$ & $x_4$ & $A_f$ & $A_r$ & $E_f$ & $E_r$ \\
    \hline
	$1$ & $1.000$  & $0.000$  & $-1.000$ & $1.000$ & 1.000e+03 & 1.801e-05 & 2.240e+03 & 5.203e+03 \\
    $2$ & $-1.000$ & $-1.000$ & $1.000$  & $0.000$ & 1.000e+00 & 4.000e+00 & 1.600e+03 & 3.680e+03 \\
    \hline
   no freezing & $x_1$ & $x_2$ & $x_3$ & $x_4$ & $A_f$ & $A_r$ & $E_f$ & $E_r$ \\
    \hline
    $1$ & $1.061$  & $0.001$ & $-1.000$ & $2.887$ & 3.391e+02 & 1.665e-06 & 2.240e+03 & 2.244e+03 \\
    $2$ & $0.969$  & $0.002$ & $-1.000$ & $0.032$ & 6.640e+02 & 2.689e-01 & 6.727e+01 & 6.684e+02 \\
    \hline    
    \end{tabular}
     \caption{{Example \ref{exam:enzyme}: learned parameters with variable temperatures. Upper part: with the PPF technique; lower part: without the PPF technique. Here $(x_1, x_2, x_3, x_4)$ denotes the row vector of the matrix $\bm{V}$.}}
     \label{tab:enzyme-params-freeze-temperature}
\end{table}

}
\end{exam}

\begin{exam}[hydrogen oxidation reaction]\label{exam:H2O}
In this example, we consider a model for hydrogen oxidation reaction where six species \ch{H2} (hydrogen), \ch{O2} (oxygen), \ch{H2O} (water), \ch{H}, \ch{O}, \ch{OH} (radicals) are involved in six steps in a closed system under constant volume and temperature \cite{gorban2005invariant,chiavazzo2008quasi}:
\begin{subequations}\label{eq:reaction-H2O}
\begin{align}
	\ch{H2 		&<=>[$k\sb{1}\sp{+}$][$k\sb{1}\sp{-}$] 2 H} \\
	\ch{O2 		&<=>[$k\sb{2}\sp{+}$][$k\sb{2}\sp{-}$] 2 O} \\
	\ch{H2O 	&<=>[$k\sb{3}\sp{+}$][$k\sb{3}\sp{-}$] H + OH} \\
	\ch{H2 + O 	&<=>[$k\sb{4}\sp{+}$][$k\sb{4}\sp{-}$] H + OH} \\
	\ch{O2 + H 	&<=>[$k\sb{5}\sp{+}$][$k\sb{5}\sp{-}$] O + OH} \\
	\ch{H2 + O 	&<=>[$k\sb{6}\sp{+}$][$k\sb{6}\sp{-}$] H2O}
\end{align}
\end{subequations}
with the reaction rates $k_1^+=2$, $k_2^+=k_3^+=1$, $k_4^+=k_5^+=1\times10^3$, $k_1^+=1\times10^2$, $k_1^- = 2.16\times10^2$, $k_2^- = 3.375\times10^2$, $k_3^- = 1.4\times10^3$, $k_4^- = 1.08\times10^4$, $k_5^- = 3.375\times10^4$, $k_6^- = 7.714285714285716\times10^{-1}$. 
The system \eqref{eq:reaction-H2O} corresponds to the simplified picture of this chemical process and the reaction rates reflect only orders of magnitude for relevant real-word systems. 
{The magnitude of the reaction rates vary from $10^{-1}$ to $10^4$, which leads to the multiscale phenomena.} {This reaction network has much more reactions and is more realistic than the first two test cases.}

We first compare the performance of our algorithm with and without the PPF treatment. The history of the training and the validation error is shown in Figure \ref{fig:H2O-loss}. Again, we observe that this technique greatly reduces the training and validation errors. The learned parameters are listed in Table \ref{tab:h2O-params-freeze}. The algorithm can generate the correct result with this technique. On the other hand, without using this technique, the phenomenon of the opposite signs observed in Table \ref{tab:hypothetical-params-freeze} also appears.
\begin{figure}
    \centering
    \includegraphics[width=0.5\textwidth]{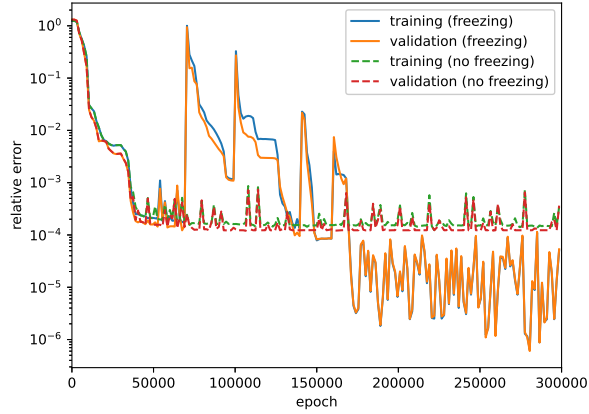}
    \caption{Example \ref{exam:H2O}: the history of the relative error for the training data and the verification data. Solid line: the method with the PPF technique; dashed line: the method without the PPF technique.}
    \label{fig:H2O-loss}
\end{figure}

\begin{table}[htbp]
  \centering
 \begin{tabular}{c|c|c|c|c|c|c|c|c}
    \hline
   freezing & $x_1$ & $x_2$ & $x_3$ & $x_4$ & $x_5$ & $x_6$ & $k_f$ & $k_r$\\
    \hline
    $1$ 	& $0.000$  & $1.000$ & $0.000$ & $1.000$ & $-1.000$ & $-1.000$ 
    & 3.375e+04 & 1.000e+03 \\
    $2$ 	& $1.000$  & $0.000$  & $0.000$ & $-1.000$ & $1.000$ & $-1.000$ 
    & 1.080e+04 & 1.000e+03 \\
    $3$ 	& $0.000$  & $0.000$  & $1.000$ & $-1.000$ & $0.000$ & $-1.000$ 
    & 1.400e+03 & 1.000e+00 \\
    $4$ 	& $0.000$  & $1.000$  & $0.000$ & $0.000$ & $-2.000$ & $0.000$ 
    & 3.375e+02 & 1.000e+00 \\
    $5$ 	& $1.000$  & $0.000$  & $0.000$ & $-2.000$ & $0.000$ & $0.000$ 
    & 2.160e+02 & 2.000e+00 \\
    $6$ 	& $-1.000$  & $0.000$  & $1.000$ & $0.000$ & $-1.000$ & $0.000$ 
    & 1.000e+02 & 7.714e$-$01 \\
    \hline
   no freezing & $x_1$ & $x_2$ & $x_3$ & $x_4$ & $x_5$ & $x_6$ & $k_f$ & $k_r$\\
    \hline
 1  & $0.000 $  &  $0.938$  &  $0.000$  &  $ 0.923$  &  $-1.000$  &  $-1.001$  & 
 	1.971e+04 & 5.512e+02 \\
 2  & $0.000 $  &  $1.087$  &  $0.000$  &  $ 1.108$  &  $-0.999$  &  $-0.999$  &
 	1.407e+04 & 4.488e+02 \\
 3  & $0.885 $  &  $0.000$  &  $0.115$  &  $-1.000$  &  $ 0.885$  &  $-1.000$  &
 	1.217e+04 & 2.112e+01 \\
 4 & $-1.004$   &  $0.000$  &  $0.087$  &  $ 0.910$  &  $-1.008$  &  $ 0.913$  &
    1.077e+03 & 2.926e+01 \\
 5  & $0.000 $  &  $0.996$  &  $0.008$  &  $ 0.000$  &  $-1.997$  &  $ 0.000$  &
    3.379e+02 & 8.912e$-$01 \\
 6  & $0.987 $  &  $0.000$  &  $0.007$  &  $-1.984$  &  $ 0.004$  &  $ 0.004$  &
    2.170e+02 & 3.377e+00 \\
    \hline    
    \end{tabular}
     \caption{Example \ref{exam:H2O}: learned parameters. Upper part: with the PPF technique; lower part: without the PPF technique. Here $(x_1, x_2, x_3, x_4, x_5, x_6)$ denotes the row vector of the matrix $\bm{V}$.}
     \label{tab:h2O-params-freeze}
\end{table}

We also test the performance of the algorithm with Gaussian noise. The algorithm can get the correct prediction of the stoichiometric coefficients with the noise level $10^{-4}$ and $10^{-3}$. The learned reaction rates with noise are shown in Table \ref{tab:h2O-params-noise-small}. The relative errors for reaction rates are typically less than the order of $10^{-2}$ for $10^{-3}$ noise and $10^{-3}$ for $10^{-4}$ noise.
\begin{table}[htbp]
  \centering
 \begin{tabular}{c|c|c|c|c}
    \hline
   	noise $10^{-3}$ & $k_f$ & relative error & $k_r$ & relative error \\
    \hline
	1 &  3.375e+04  &  5.706e$-$05  &  1.002e+03  &  2.097e$-$03  \\	 
	2 &  1.080e+04  &  2.789e$-$04  &  1.001e+03  &  1.374e$-$03  \\
	3 &  1.399e+03  &  4.413e$-$04  &  9.631e$-$01  &  3.836e$-$02  \\
	4 &  3.399e+02  &  7.103e$-$03  &  9.235e$-$01  &  8.278e$-$02  \\
	5 &  2.161e+02  &  6.927e$-$04  &  2.130e+00  &  6.107e$-$02  \\
	6 &  9.764e+01  &  2.417e$-$02  &  8.047e$-$01  &  4.137e$-$02  \\   	 
    \hline
    noise $10^{-4}$ & $k_f$ & relative error & $k_r$ & relative error \\
    \hline
	1 &  3.375e+04     &  6.482e$-$06  &  1.000e+03     & 2.099e$-$04 \\ 
	2 &  1.080e+04     &  2.803e$-$05  &  1.000e+03     & 1.379e$-$04 \\   
	3 &  1.400e+03     &  4.491e$-$05  &  9.963e$-$01     & 3.707e$-$03 \\
	4 &  3.377e+02     &  7.161e$-$04  &  9.923e$-$01     & 7.717e$-$03 \\
	5 &  2.160e+02     &  6.965e$-$05  &  2.013e+00     & 6.469e$-$03 \\
	6 &  9.976e+01     &  2.366e$-$03  &  7.748e$-$01     & 4.294e$-$03 \\
    \hline    
    \end{tabular}
     \caption{Example \ref{exam:H2O}: learned reaction rates with noise.}
     \label{tab:h2O-params-noise-small}
\end{table}

Moreover, we plot the evolution of the concentrations of the six species with the noise level $10^{-3}$ in Figure \ref{fig:H2O-reaction-time-exact}. We observe a good agreement of the solution generated by our learned model and the exact solution.  We also measure the prediction errors of the learned model at 100 uniformly points in the time interval $[0, 10]$. The prediction errors are $1.953\times10^{-6}$ with zero noise, $9.152\times10^{-4}$ with noise level $10^{-4}$ and $8.710\times10^{-4}$ with noise level $10^{-3}$.
\begin{figure}
    \centering
    \includegraphics[width=0.5\textwidth]{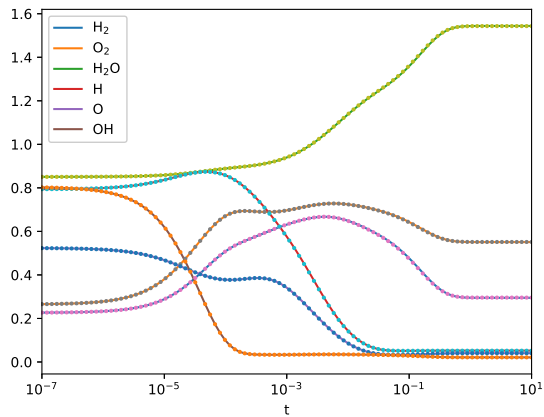}
    \caption{Example \ref{exam:H2O}: the evolution of the concentration of the 6 species in the hydrogen oxidation reaction problem obtained by solving the original ODEs \eqref{eq:reaction-H2O} and our learned ODEs. noise level $10^{-3}$.}
    \label{fig:H2O-reaction-time-exact}
\end{figure}

\end{exam}

{

\begin{exam}[extended Zeldovich mechanism]\label{exam:Zeldovich}
In this example, we test our algorithm on the extended Zeldovich mechanism, which is a chemical mechanism describing the oxidation of nitrogen and \ch{NOx} formation \cite{zeldovich1985mathematical}. {Similar to Example \ref{exam:H2O}, this is another realistic test case.} The reaction mechanisms read as
\begin{subequations}\label{eq:reaction-NOX}
\begin{align}
	\ch{N2 + O  &<=>[$k\sb{1}\sp{+}$][$k\sb{1}\sp{-}$] NO + N} \\
	\ch{N + O2 	&<=>[$k\sb{2}\sp{+}$][$k\sb{2}\sp{-}$] NO + O} \\
	\ch{N + OH 	&<=>[$k\sb{3}\sp{+}$][$k\sb{3}\sp{-}$] NO + H}
\end{align}
\end{subequations}
and the reaction rates are given by the Arrhenius law \cite{hanson1984survey}:
\begin{equation}
\begin{aligned}
	k_1^+ &= 1.8\times10^{11} \exp(-38370/T), \quad k_1^- = 3.8\times10^{10} \exp(-425/T), \\
	k_2^+ &= 1.8\times10^{7} \exp(-4680/T), \quad k_2^- = 3.8\times10^{6} \exp(-20820/T), \\
	k_3^- &= 7.1\times10^{10} \exp(-450/T), \quad k_3^- = 1.7\times10^{11} \exp(-24560/T),
\end{aligned}
\end{equation}
with $T$ the temperature.

In the numerical test, we fix the temperature to be $T=3000$, which is a reasonable temperature in real applications \cite{hanson1984survey}. At this temperature, the reaction rates are 
\begin{equation}
\begin{aligned}
	& k_1^+ = 5.019\times10^5, \quad k_2^+ = 3.782\times10^6, \quad k_3^+ = 6.111\times10^{10}, \\
	& k_1^- = 3.298\times10^{10}, \quad k_2^- = 3.679\times10^3, \quad k_3^- = 4.732\times10^7.
\end{aligned}
\end{equation}  
Then, we follow the same procedure in the previous examples to generate the data and execute the algorithm to discover the stoichiometric coefficients and the reaction rates. Again, the algorithm with the PPF treatment can predict the correct result, which is shown in Table \ref{tab:Zeldovich-params-freeze}. We observe that the accurate reaction rates are obtained.
\begin{table}[htbp]
  \centering
 \begin{tabular}{c|c|c|c|c|c|c|c|c|c}
    \hline
   freezing & $x_1$ & $x_2$ & $x_3$ & $x_4$ & $x_5$ & $x_6$ & $x_7$ & $k_f$ & $k_r$\\
    \hline
    $1$ 	& $0.000$  & $0.000$ & $1.000$ & $-1.000$ & $0.000$ & $-1.000$ & $1.000$ 
    & 6.111e+10 & 4.732e+07 \\
    $2$ 	& $1.000$  & $1.000$  & $-1.000$ & $-1.000$ & $0.000$ & $0.000$ & $0.000$ 
    & 3.298e+10 & 5.019e+05 \\
    $3$ 	& $0.000$  & $1.000$  & $1.000$ & $-1.000$ & $-1.000$ & $0.000$ & $0.000$
    & 3.782+06 & 3.931e+03 \\
    \hline
    \end{tabular}
    \caption{{ Example \ref{exam:Zeldovich}: learned parameters with the PPF technique. Here $(x_1, x_2, x_3, x_4, x_5, x_6, x_7)$ denotes the row vector of the matrix $\bm{V}$.}}
     \label{tab:Zeldovich-params-freeze}
\end{table}

\end{exam}
}

{

\begin{exam}[simplified GRI-3.0 mechanism]\label{exam:GRI3}
In this example, we test our algorithm on the simplified GRI-3.0 mechanism, which is a chemical mechanism describing the methane oxidation \cite{2001Augmented}. {This is the most complicated reaction system tested in the paper.} The mechanism includes 16 species with 12 reactions and reads as
\begin{subequations}\label{eq:reaction-NOX}
\begin{align}
	\ch{CH_4 + H & ->[$k\sb{1}$] CH_3 + H_2} \\
	\ch{CH_2O + H_2 & ->[$k\sb{2}$] CH_3 + OH} \\
	\ch{CH_2O & ->[$k\sb{3}$] CO + H_2} \\
	\ch{C_2H_6 & ->[$k\sb{4}$] C_2H_4 + H_2} \\
	\ch{C_2H_4 + OH & ->[$k\sb{5}$] CH_3 + CO + H_2} \\
	\ch{2 CO + H_2 & ->[$k\sb{6}$] C_2H_2 + O_2} \\
	\ch{CO + OH + H & ->[$k\sb{7}$] CO_2 + H_2} \\
	\ch{H + OH & ->[$k\sb{8}$] H_2O} \\
	\ch{2 H + 2 OH & ->[$k\sb{9}$] 2 H_2 + O_2} \\
    \ch{H_2 & ->[$k\sb{10}$] 2 H} \\
    \ch{H_2 + O_2 & ->[$k\sb{11}$] HO_2 + H} \\
    \ch{H_2O_2 + H & ->[$k\sb{12}$] H_2 + HO_2} 
\end{align}
\end{subequations}
The reaction rates are given in \cite{2001Augmented}, which are derived from the reaction rates of the standard GRI-3.0 Mech \cite{GRI}. We compute the reaction rates with the temperature $T=3000$ and list them in Table 4.8. Here, the reaction rates are normalized such that the smallest one is of order 1.
\begin{table}[htbp]
  \centering
 \begin{tabular}{c|c|c|c|c|c}
    \hline
    $k_1$ & $k_2$ & $k_3$ & $k_4$ & $k_5$ & $k_6$ \\
    \hline
    5.088e+00 & 1.891e+00 & 2.607e+00 & 6.268e+00 & 5.446e+00 & 1.283e+01 \\
    \hline
    \hline
    $k_7$ & $k_8$ & $k_9$ & $k_{10}$ & $k_{11}$ & $k_{12}$ \\
    \hline
    1.349e+00 & 5.264e+03 & 3.268e+01 & 4.873e+03 & 2.978e+02 & 5.227e+03\\
    \hline
    \end{tabular}
    \label{tab:GRI3-params-exact}
    \caption{{Example \ref{exam:GRI3}: reaction rates in simplified GRI-3.0 Mech}}
\end{table}

Note that all the reactions in \eqref{eq:reaction-NOX} are not reversible. Here, we apply exactly the same algorithm to this situation, similar to Example \ref{exam:enzyme}. 
To illustrate the advantage of the PPF technique, we first compare the performance of the algorithm with and without this technique. The history of the training and validation errors is shown in Figure \ref{fig:GRI3-loss-freeze}. The relative error stays around $10^{-3}$ without this technique, and decreases to $10^{-6}$ after applying this technique.
\begin{figure}
    \centering
    \includegraphics[width=0.5\textwidth]{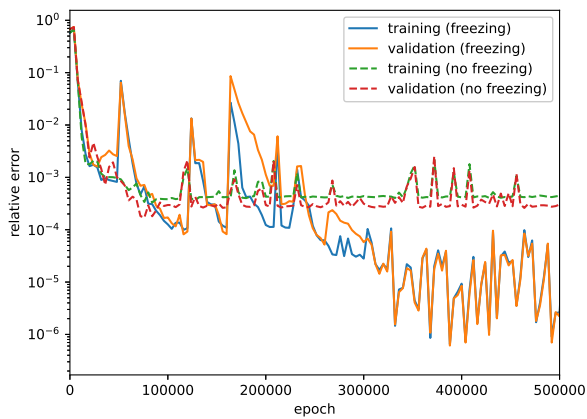}
    \caption{{Example \ref{exam:GRI3}: the history of the relative error for the training data and the verification data. Solid line: the method with the PPF technique; dashed line: the method without the PPF technique.}}
    \label{fig:GRI3-loss-freeze}
\end{figure}

\begin{table}[htbp]
  \centering
 \begin{tabular}{c|c|c|c|c|c}
    \hline
    $k^+_1$ & $k^+_2$ & $k^+_3$ & $k^+_4$ & $k^+_5$ & $k^+_6$ \\
    \hline
    5.088e+00 & 1.891e+00 & 2.607e+00 & 6.268e+00 & 5.446e+00 & 1.283e+01 \\
    \hline
    \hline
    $k^+_7$ & $k^+_8$ & $k^+_9$ & $k^+_{10}$ & $k^+_{11}$ & $k^+_{12}$ \\
    \hline
    1.349e+00 & 5.264e+03 & 3.268e+01 & 4.873e+03 & 2.978e+02 & 5.227e+03\\
    \hline
    \hline
    $k^-_1$ & $k^-_2$ & $k^-_3$ & $k^-_4$ & $k^-_5$ & $k^-_6$ \\
    \hline
    2.546e-04 & 1.695e-04 & 1.091e-04 & 1.751e-04 & 1.103e-04 & 9.052e-06 \\
    \hline
    \hline
    $k^-_7$ & $k^-_8$ & $k^-_9$ & $k^-_{10}$ & $k^-_{11}$ & $k^-_{12}$ \\
    \hline
    8.462e-05 & 1.146e-05 & 5.472e-04 & 2.625e-07 & 8.566e-07 & 3.863e-04\\
    \hline    
    \end{tabular}
    \label{tab:GRI3-params-freeze}
    \caption{{Example \ref{exam:GRI3}: learned reaction rates in simplified GRI-3.0 Mech. Upper part: reaction rates in the forward reaction; lower part: reaction rates in the reverse reaction.}}
\end{table}
We also list the learned parameters with the PPF technique in Table 4.9. 
Here, the learned stoichiometric coefficients are the same with the true coefficients in \eqref{eq:reaction-NOX} and they are omitted here. The upper part of the table is the learned rates in the forward reactions with the PPF technique, which agrees well with the ground truth in Table 4.8.
The learned rates in the reverse reactions are in the magnitude of $10^{-7}$ to $10^{-4}$. It then can be inferred that the system can be well described using only forward reactions. By contrast, the algorithm without imposing this technique could not generate the correct result and we omit the results here.

\end{exam}
}

\section{Conclusion}\label{sec:conclusion}

In this paper, we propose a data-driven method to discover multiscale chemical reactions governed by the law of mass action. The method mainly contains two novel points.
First, we use a single matrix to represent the stoichiometric coefficients for both the reactants and products in a system without catalysis reactions. 
The negative entries in the matrix denote the stoichiometric coefficients for the reactants and the positive ones for the products.
Second, by considering a multiscale nonlinear regression problem, we find that the conventional optimization methods usually get stuck in the local minima and could not find the true solution. To escape from the local minima, we propose a PPF technique. Notice that the stoichiometric coefficients are integers. In the training process, if the loss function stops to decrease, the stoichiometric coefficients which are close to integers are rounded and then frozen afterwards. With such a treatment, the stoichiometric coefficients are gradually determined, the dimension of the searching space is reduced in the training process, and eventually the global mimina can be obtained. Several numerical experiments including the classical Michaelis–Menten kinetics, the hydrogen oxidation reactions and simplified GRI-3.0 mechanism verify the validity of our algorithm in learning the multiscale chemical reactions.

There are still some problems to be addressed in order to develop a robust and general framework for discovering multiscale chemical reactions from data. We shall highlight some of the challenges that could guide future advances. First, it is interesting to generalize the PPF technique to the catalysis reactions. 
{ 
Second, the number of species $n_s$ cannot be determined from our algorithm. In principle, to infer the unknown chemical reaction systems, we should have the concentration time series data for all the species. Our algorithm cannot treat the problem when the concentrations for some partial species are unknown. This difficulty may be overcomed by combining the current algorithm with the Neural ODE approach in \cite{ji2020autonomous}.
The third challenge is that for very complex reaction networks with large number of reactions (hundreds or thousands), our algorithm may not always find out the correct solution. New ideas are needed at this point. }

\bibliographystyle{abbrv}
\bibliography{chemical_ref}

\end{document}